\documentclass[aps,prl,twocolumn,preprintnumbers,10pt,showpacs]{revtex4-1}
\usepackage{amsmath,bm,amssymb}

\usepackage{graphicx}

\newcommand{\dsigma}{{\rm d}\sigma}
\def\gapprox{\lower .7ex\hbox{$\;\stackrel{\textstyle >}{\sim}\;$}}

\allowdisplaybreaks

\begin{document}

\preprint{IPPP/14/64, ZU-TH 27/14}

\title{Precise QCD predictions for the production of Higgs+jet final states}

\author{X.\ Chen$^a$, T.\ Gehrmann$^b$, E.W.N.\ Glover$^a$, M.\ Jaquier$^b$}

\affiliation{$^a$ Institute for Particle Physics Phenomenology, Department of Physics, University of Durham, Durham, DH1 3LE, UK\\
$^b$ Department of Physics, University of
  Z\"urich, CH-8057 Z\"urich, Switzerland}


\begin{abstract}
  We compute the cross section 
  and differential distributions for the production of a Standard Model Higgs boson in 
  association with a hadronic jet to next-to-next-to-leading order in 
 quantum chromodynamics (QCD).
  In Higgs boson studies at the  LHC, final states containing one jet are a dominant 
  contribution to the total event rate, and their understanding is crucial for improved determinations 
 of the Higgs boson properties. 
 We observe substantial higher order corrections to 
  transverse momentum spectra and rapidity distributions in Higgs-plus-one-jet 
 final states. Their inclusion stabilises the residual theoretical uncertainty of the predictions around
 9\%, thereby providing important input to precision studies of the Higgs boson.  
 \end{abstract}

\maketitle

After the initial observation of the Higgs boson at the CERN LHC~\cite{higgsLHC}, several 
measurements of its properties (production and decay modes, spin) have provided 
the first evidence for its nature~\cite{higgspropsLHC}, which is found to be largely in agreement 
(still with substantial error margins) with the 
expectations of the Standard Model. Precision studies of Higgs boson properties will 
be among the primary objectives of the upcoming run at the LHC, allows searches for 
possible small deviations  from the
Standard Model formulation of the Higgs mechanism of electroweak symmetry breaking. 
The interpretation of these precision data relies on a close interplay 
between experimental measurements and theoretical calculations for 
Higgs boson signal and background processes. 

To obtain reliable theoretical predictions for hadron collider observables, corrections from 
higher order processes in QCD need to be accounted for. For the most important 
Higgs boson production processes, impressive progress has been made in recent years, 
such that gluon 
fusion~\cite{ggHnnlo} and associated production~\cite{vhnnlo} are described 
fully exclusively to next-to-next-to-leading 
order (NNLO) in QCD, and vector-boson-fusion~\cite{vbfnlo} and associated production 
with top quarks~\cite{ttHnlo} to next-to-leading order (NLO).  
For the gluon fusion process, which is the largest  contribution to Higgs production 
at the LHC, NLO corrections 
have also been derived for Higgs boson production in association with up to 
three jets~\cite{ggH1jnlo,ggH2jnlo,ggH3jnlo}. Recently, the first steps towards the fourth-order QCD corrections 
(N$^3$LO) have been taken~\cite{ggHnew}.

The number of jets produced in association with a Higgs boson candidate is 
a very important discriminator between different production modes, and is utilised in the 
optimization of signal-to-background ratios, for example through the application of jet 
vetoes~\cite{veto1}. For many Higgs boson studies, final states 
with $H+0$~jets and $H+1$~jet contribute roughly equal amounts to the total 
cross section. A comparable theoretical precision for both processes in the 
dominant gluon fusion production mode is 
therefore mandatory for precision studies and to resolve potential correlations between 
both samples~\cite{hcorr1,hcorr2}. 
While 
the $H+0$~jets  process is 
described to NNLO~\cite{ggHnnlo} accuracy, 
 $H+1$~jet production is known fully differentially only to NLO~\cite{ggH1jnlo}. 
A first step towards the 
 NNLO corrections for this process has been taken in~\cite{caola} 
with the purely gluonic contribution to the total $H+1$~jet cross section. This work 
has highlighted that NNLO contributions to $H+1$~jet final states are of substantial 
numerical magnitude~\cite{caola} and clearly called for a more differential 
description. In this letter, we report on a new 
calculation of the gluonic NNLO contributions to $H+1$~jet production in gluon fusion, 
carried out in the form of a parton-level event generator that provides the corrections in 
a fully differential form, including the Higgs decay to two photons.  An extension to 
different Higgs boson decay modes is feasible. As pointed out in Ref.~\cite{caola}, the gluonic process dominates over the other subprocesses including the potentially large quark-gluon channel.   We note that the techniques employed here can be
applied to these other contributions.

\begin{figure}[t]
(a)\includegraphics[width=2.2cm]{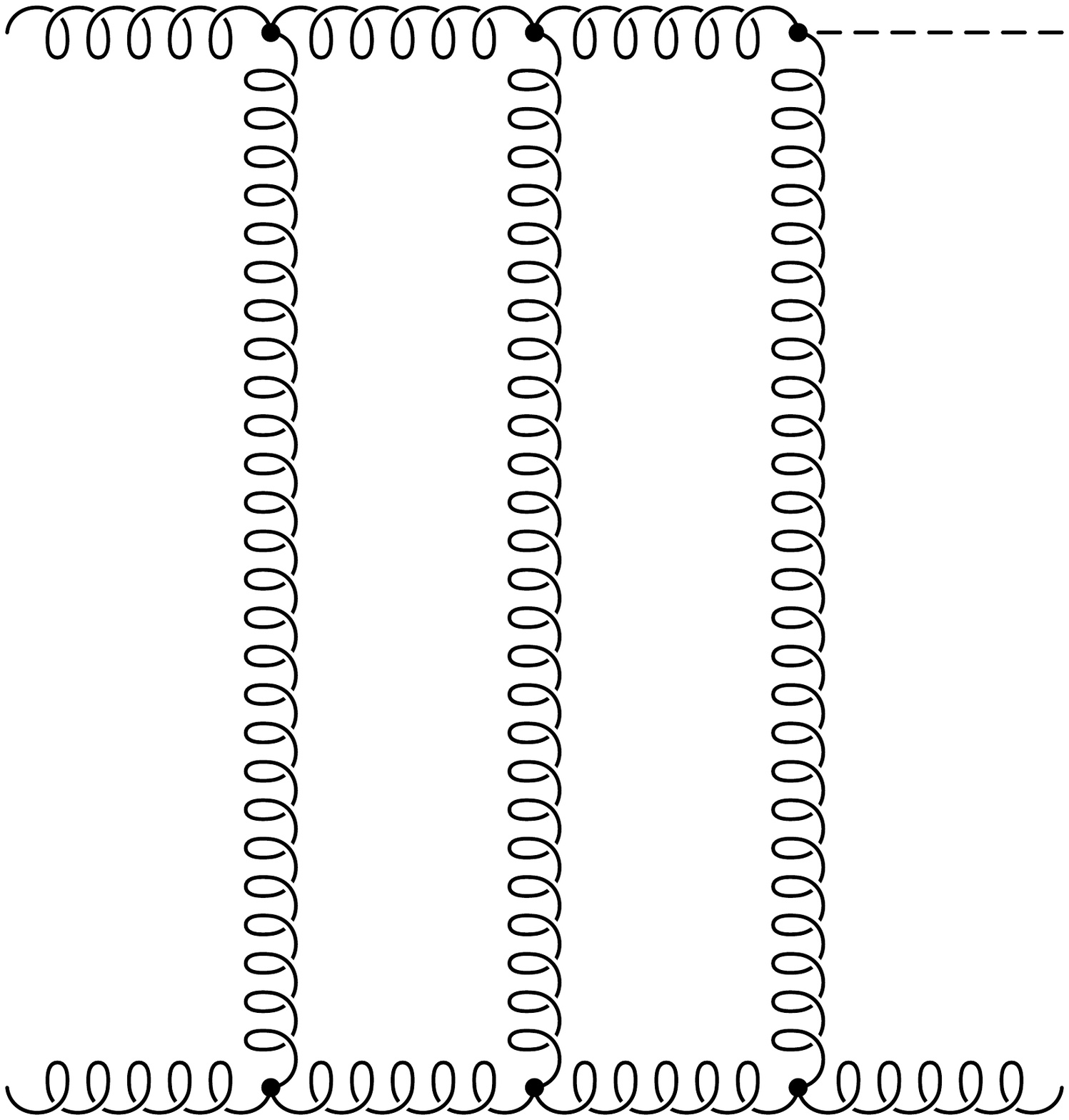}
(b)\includegraphics[width=2.2cm]{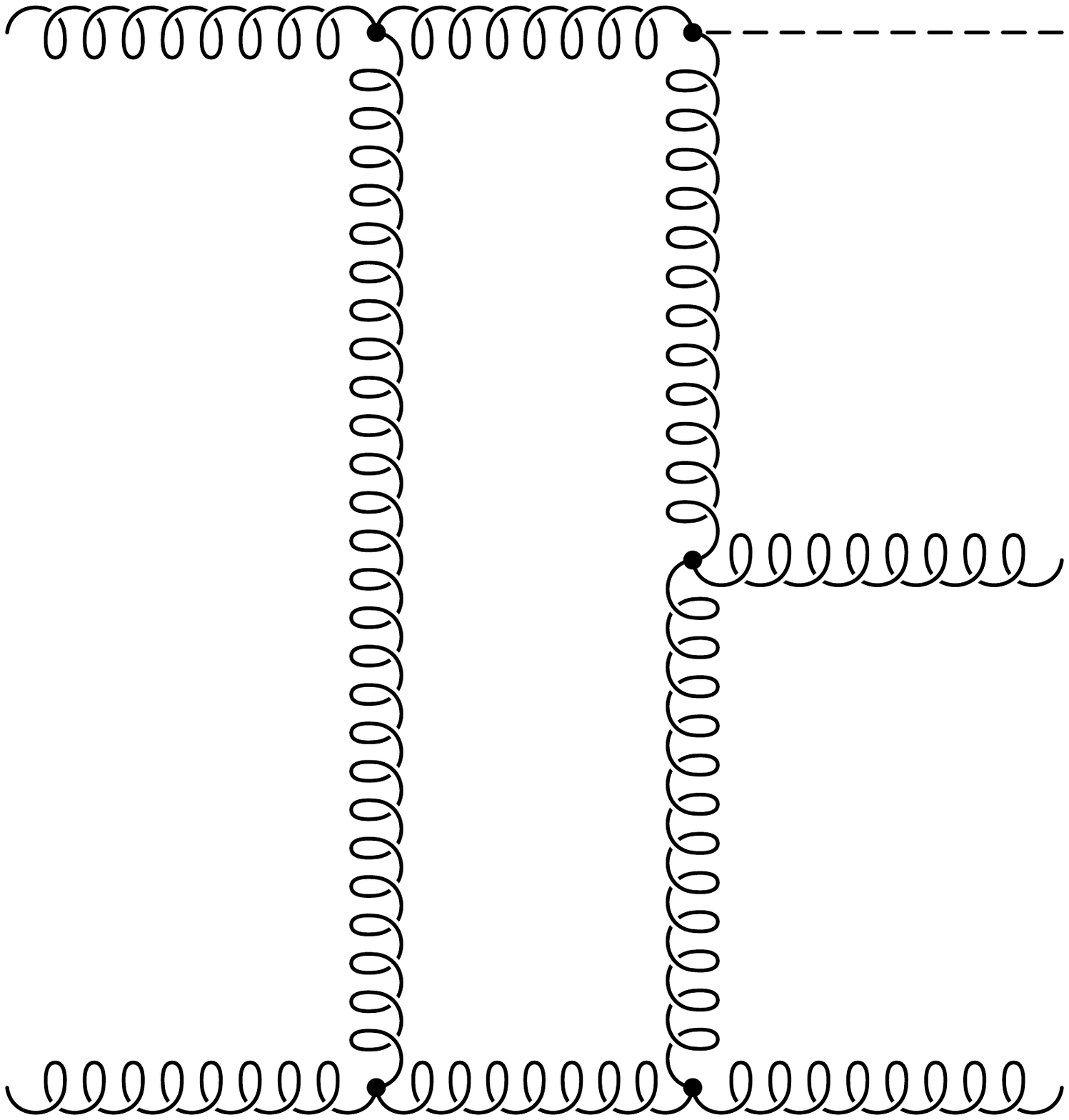}
(c)\includegraphics[width=2.2cm]{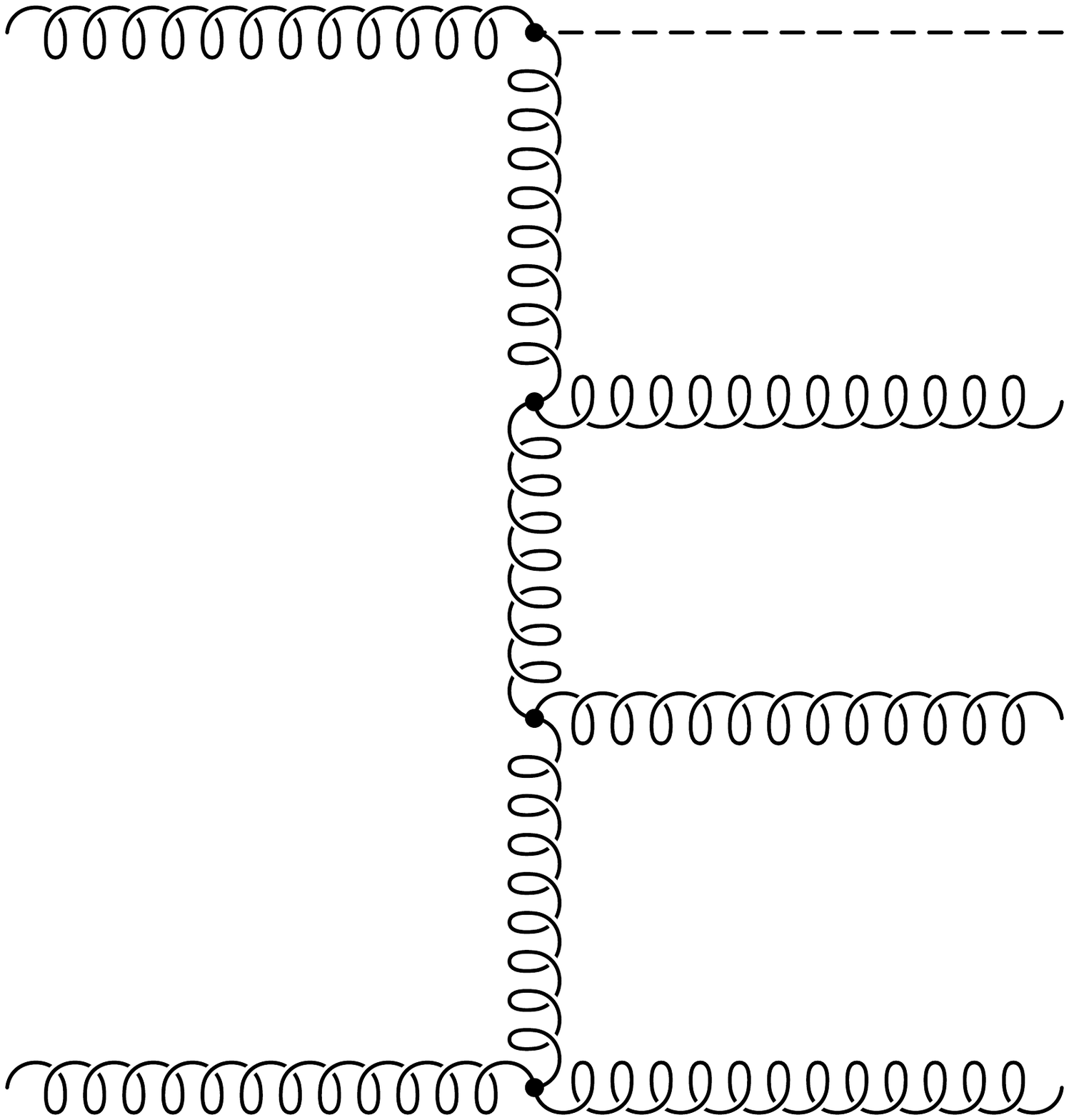}
\caption{Representative Feynman diagrams in the effective theory for  
(a) two-loop Higgs-plus-three-gluon amplitudes (b)
one-loop Higgs-plus-four-gluon amplitudes and (c) 
tree-level Higgs-plus-five-gluon amplitudes.\label{fig:FD}}
\end{figure}

The NNLO corrections to Higgs+jet production 
in hadronic collisions receive contributions from the three types of parton-level
processes: (a) the two-loop corrections to Higgs-plus-three-parton processes~\cite{h3g2l}, (b)
the one-loop 
corrections to Higgs-plus-four-parton processes~\cite{h4g1l} and (c) 
the tree-level Higgs-plus-five-parton 
processes~\cite{h5g0l}. Figure~\ref{fig:FD} shows representative Feynman diagrams for each of the gluonic processes.
The effective interaction between gluons and the Higgs boson is mediated by top quarks and is valid in the $m_t \to \infty$ limit. 
The ultraviolet renormalised matrix elements for these processes are integrated 
over the final state phase space appropriate to Higgs+jet final states. All three types of contributions 
are infrared-divergent, only their 
sum is finite. 
While infrared divergences from the virtual corrections are explicit in the 
one-loop and two-loop matrix elements, divergences from unresolved real radiation become 
explicit only after phase space integration. The divergences are usually regulated dimensionally, and 
different methods have been used for their extraction from the real radiation 
contributions. All these methods are based on a subtraction of divergent configurations, which are 
then integrated over the phase space and added to the virtual corrections to yield a finite result: 
sector decomposition~\cite{secdec}, antenna subtraction~\cite{ourant}, $q_T$-subtraction~\cite{qtsub}
and sector-improved residue subtraction~\cite{stripper} have all been applied successfully in the calculation of NNLO corrections to exclusive processes. 

In this calculation we apply antenna subtraction, a method for the construction of 
real radiation subtraction terms based on so-called 
antenna functions, that each describe all infrared singular limits occurring in between two hard 
colour-ordered partons. For hadron-collider observables, either hard radiator can be in the 
initial or final state, and all unintegrated and integrated 
antenna functions were derived previously~\cite{hadant,gionata,monni,ritzmann}. 
The gluonic cross-section is given by,
\begin{eqnarray}
\dsigma_{gg,NNLO}&=&\int_{{\rm{d}}\Phi_{3}}\left[\dsigma_{gg,NNLO}^{RR}-\dsigma_{gg,NNLO}^S\right]
\nonumber \\
&+& \int_{{\rm{d}}\Phi_{2}}
\left[
\dsigma_{gg,NNLO}^{RV}-\dsigma_{gg,NNLO}^{T}
\right] \nonumber \\
&+&\int_{{\rm{d}}\Phi_{1}}\left[
\dsigma_{gg,NNLO}^{VV}-\dsigma_{gg,NNLO}^{U}\right],
\end{eqnarray}
where each of the square brackets is finite and well 
behaved in the infrared singular regions. 
The construction of the subtraction terms $\dsigma_{gg,NNLO}^{S,T,U}$ follows 
closely the NNLO subtraction terms for purely gluonic jet production~\cite{joao}. 

Using the antenna subtraction method to cancel infrared divergent terms 
between different channels, we have implemented all purely gluonic subprocesses 
to Higgs-plus-jet production through to
 NNLO into a parton-level event generator. With this program, we can 
compute any infrared safe observable related to $H+j$ final states to NNLO accuracy. The Higgs decay 
to two photons is included, such that realistic event selection cuts on the photons can equally be 
applied once multiple differential distributions become available.
Renormalization and factorization scales can be chosen on an event-by-event basis.

For our numerical computations, we use the NNPDF2.3 parton distribution functions~\cite{nnpdf}
with the corresponding value of $\alpha_s(M_Z)=0.118$ at NNLO, and $M_H=125~$GeV. Default 
values for the factorization and renormalization scales are $\mu_F=\mu_R=M_H$, with 
theory errors estimated from the envelope 
of a  variation to $M_H/2$ and $2\,M_H$. To compare 
with previously obtained results for the total cross section for 
purely gluonic $H+j$ production at $\sqrt{s}=8$~TeV, we use the same cuts as in~\cite{caola}: jets are reconstructed 
in the $k_T$ algorithm with $R=0.5$, and accepted if $p_T>30$~GeV.
With this, we obtain the total cross section at different  perturbative orders as
\begin{eqnarray}
\sigma_{LO} &=& 2.72^{+1.22}_{-0.78}~\mbox{ pb}\;,\nonumber \\
\sigma_{NLO} &=& 4.38^{+0.76}_{-0.74}~\mbox{ pb}\;,\nonumber \\
\sigma_{NNLO} &=& 6.34^{+0.28}_{-0.49}~\mbox{ pb}\;,
\end{eqnarray} 
in very good agreement with~\cite{caola}. 
\begin{figure}[t]
\begin{center}
(a)
\includegraphics[angle=-90,width=7cm]{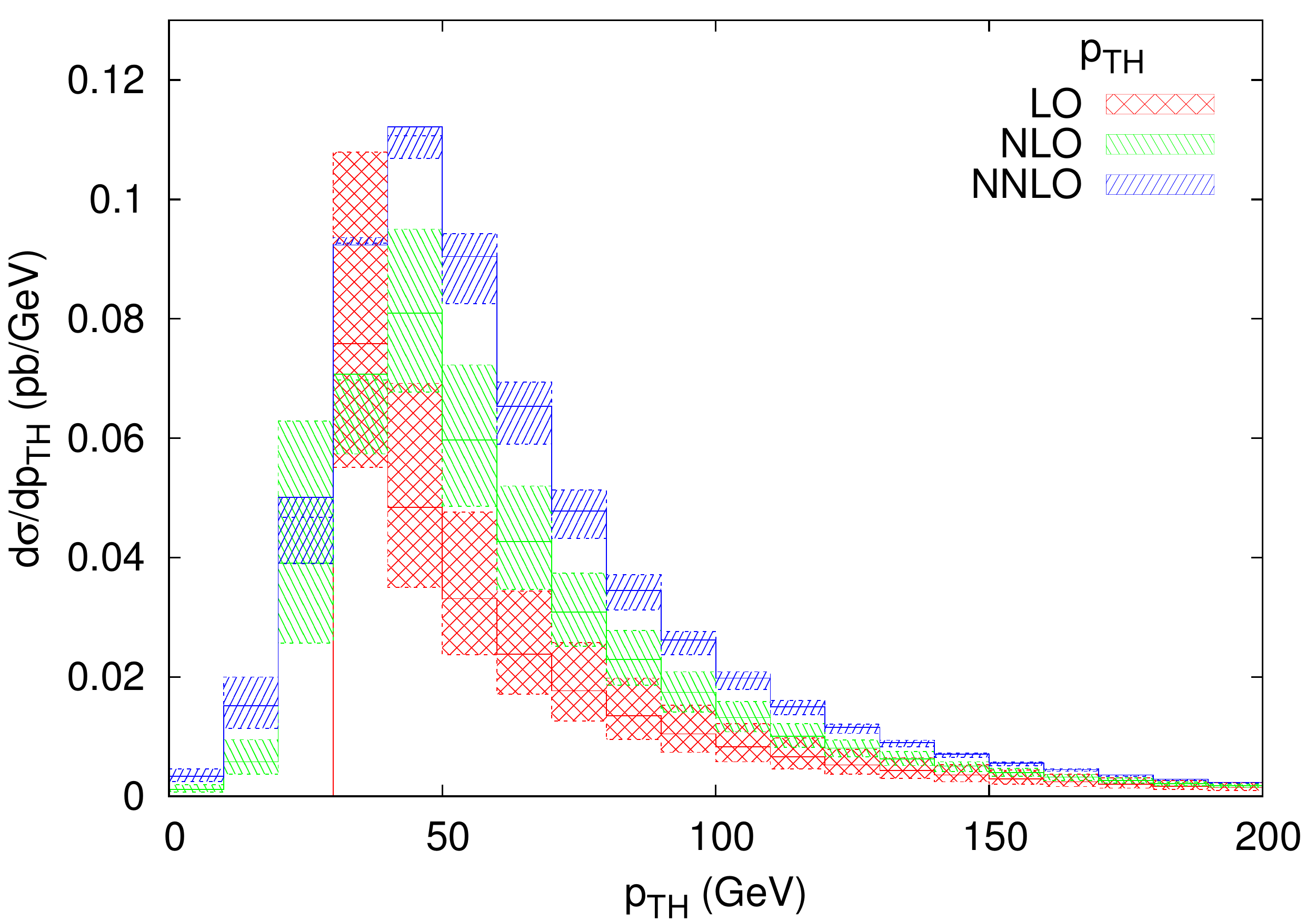}

(b)
\includegraphics[angle=-90,width=7cm]{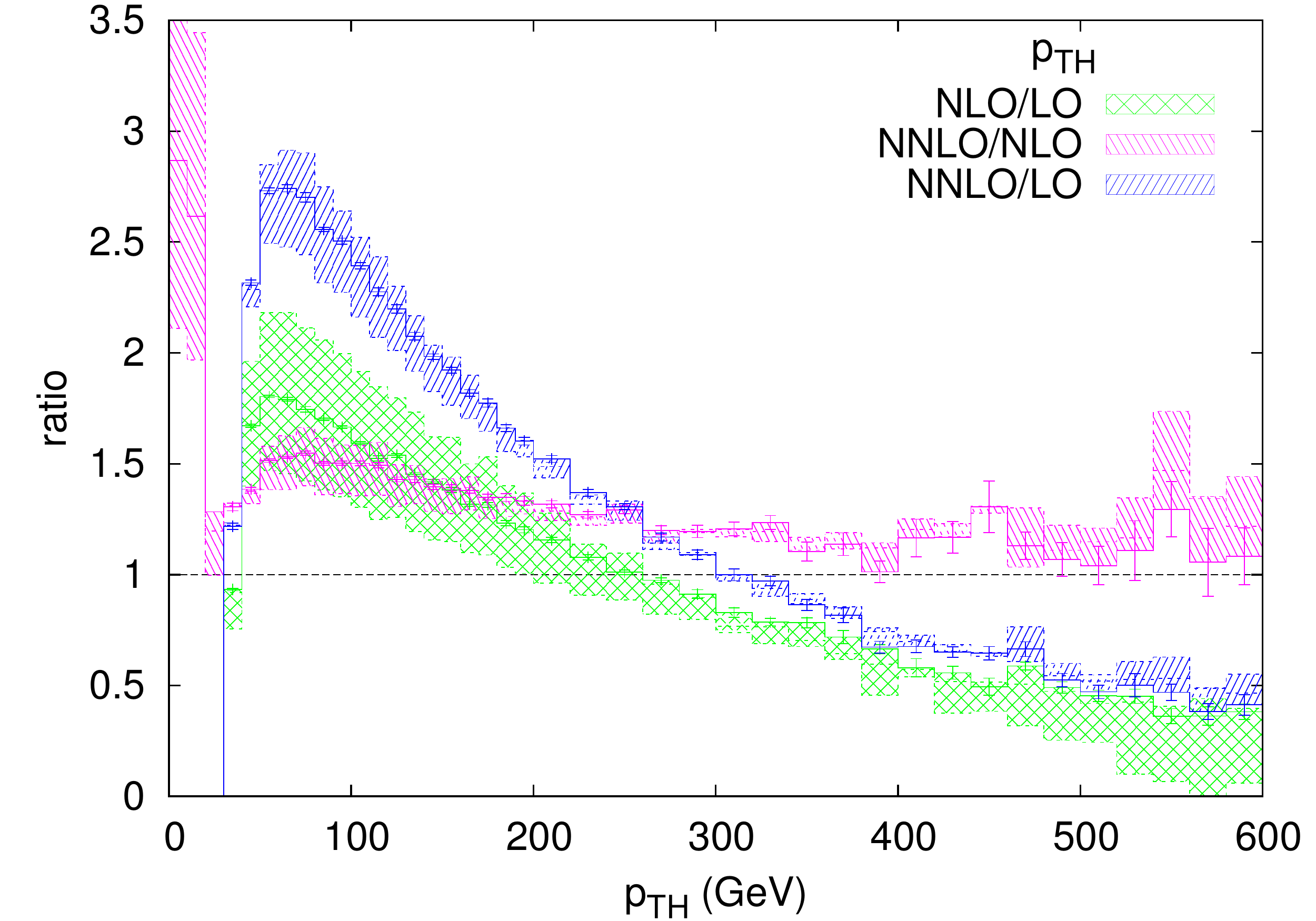}
\end{center}
\caption{(a) Transverse momentum distribution of the Higgs boson
in inclusive $H+1j$ production in $pp$ collisions with 
$\sqrt{s}=8$~TeV at LO, NLO, NNLO and (b) 
Ratios of different perturbative orders, NLO/LO, NNLO/LO and NNLO/NLO.\label{fig:hqt}}
\end{figure}

In our  kinematical distributions and ratio plots, the 
error band describes the scale variation 
envelope as described above, where the denominator in the ratio plots is  evaluated at fixed central
scale, such that the band only reflects the variation of the numerator. Error bars on the distributions 
indicate the numerical integration errors on individual bins. 
\begin{figure}[t]
\begin{center}
(a)
\includegraphics[angle=-90,width=7cm]{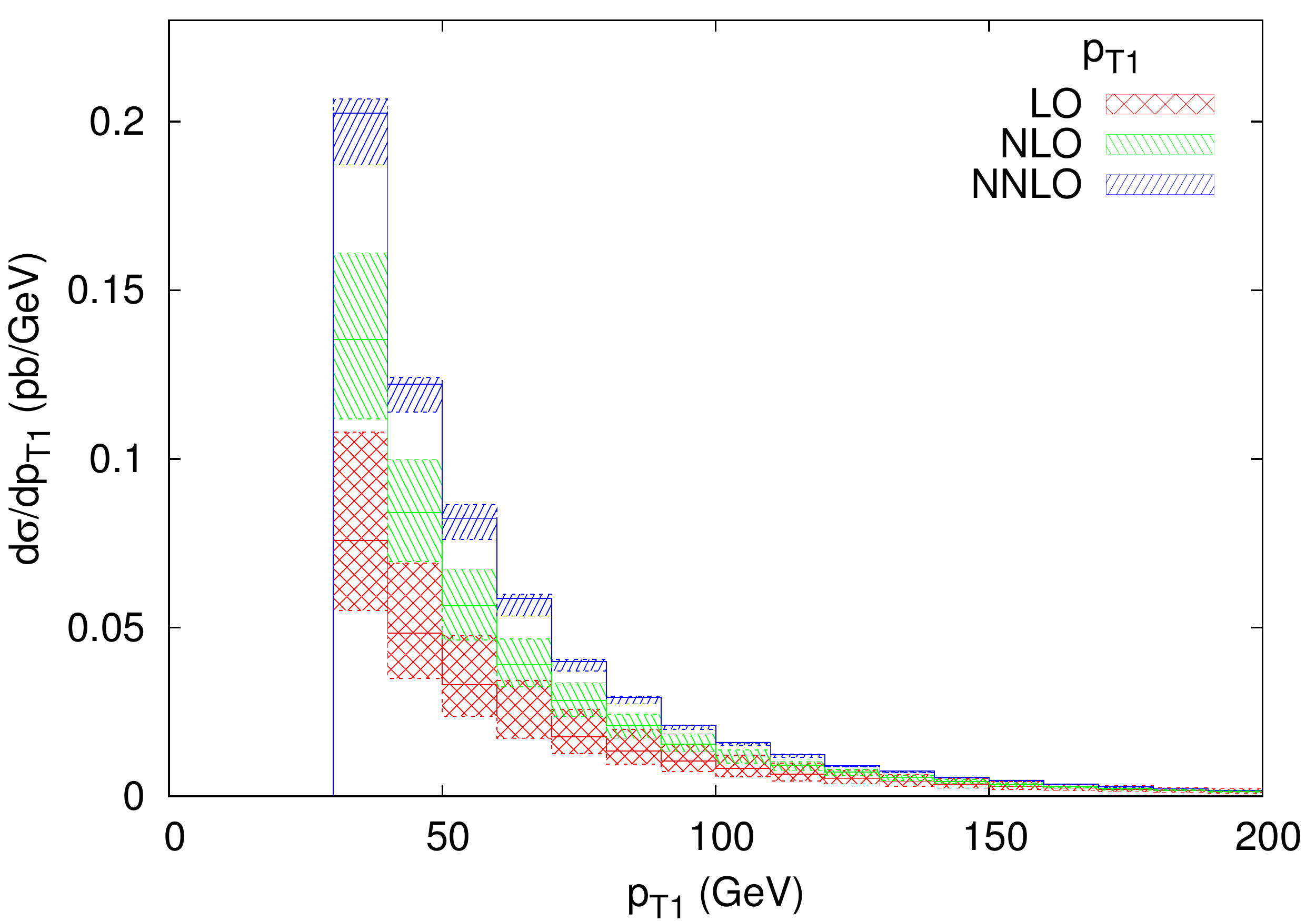}

(b)
\includegraphics[angle=-90,width=7cm]{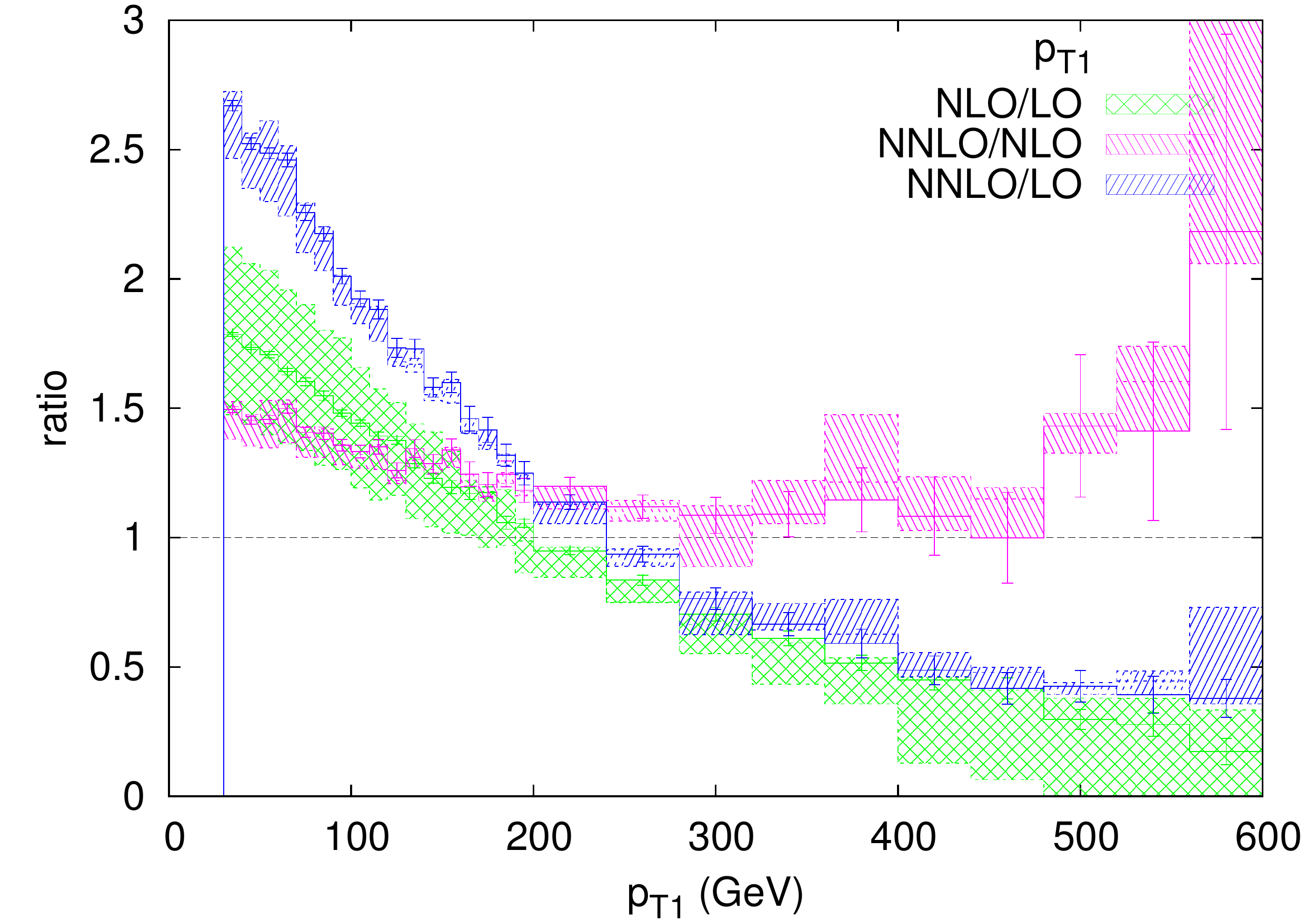}
\end{center}
\caption{(a) Transverse momentum distribution of the leading jet
in inclusive $H+1j$ production  in $pp$ collisions with 
$\sqrt{s}=8$~TeV
at LO, NLO, NNLO and (b) Ratios
of different perturbative orders, NLO/LO, NNLO/LO and NNLO/NLO.\label{fig:j1qt}}

\end{figure}

The transverse momentum distribution of the Higgs boson is particularly important for discriminating between different Higgs production modes.
A first measurement has recently been presented by 
ATLAS~\cite{atlashqt}, demonstrating the feasibility and future experimental prospects
for this observable.   It has been computed previously to NLO~\cite{ggH1jnlo}, 
combined with resummation to third logarithmic order (NNLL)~\cite{hqtresum}. 
In Figure~\ref{fig:hqt}, 
we observe that the Higgs boson transverse momentum distribution receives 
sizable NNLO corrections throughout the whole range in $p_T$, which enhance 
the NLO cross section by a quasi constant factor of about 1.4, slightly 
decreasing towards higher $p_T$. Using the same scale variations pattern as for the 
inclusive cross section above, we observe that the $p_T$ distribution of the Higgs boson has 
a residual NNLO theory uncertainty ranging between 5\% and 16\%.  
At high values of $p_T$, the effective theory 
approximation used for the coupling of the Higgs boson to gluons breaks down, since 
the large momentum transfer in the process  starts resolving the top quark loop. Consequently, one 
expects top quark mass effects for $p_T \gapprox m_t$  to be more important than the higher order corrections in 
the effective theory. 

We note that at leading order $p_{T,H}$ is 
kinematically forced to be equal to the transverse momentum of the jet, and is consequently larger 
than the transverse momentum cut on the jet. At higher orders, higher multiplicity final states are allowed
and this kinematical restriction no longer applies. These kinematical situations often lead to 
instabilities in the perturbative expansion, with large corrections at each order. 
We observe that this is not the case here: NNLO corrections 
to the Higgs boson $p_T$ distribution in $H+j$ events turn out to be moderate below the 
jet cut of $p_T=30$~GeV. 
A similar
pattern to the Higgs $p_T$ distribution, 
 is also observed for the leading jet, Figure~\ref{fig:j1qt}, which displays a slightly smaller scale
 uncertainty amounting up to 12\%, and displays rising NNLO corrections for very large values of 
 $p_T$, again likely beyond the applicability of the effective theory approximation.
\begin{figure}[t]
\begin{center}
(a)
\includegraphics[angle=-90,width=7cm]{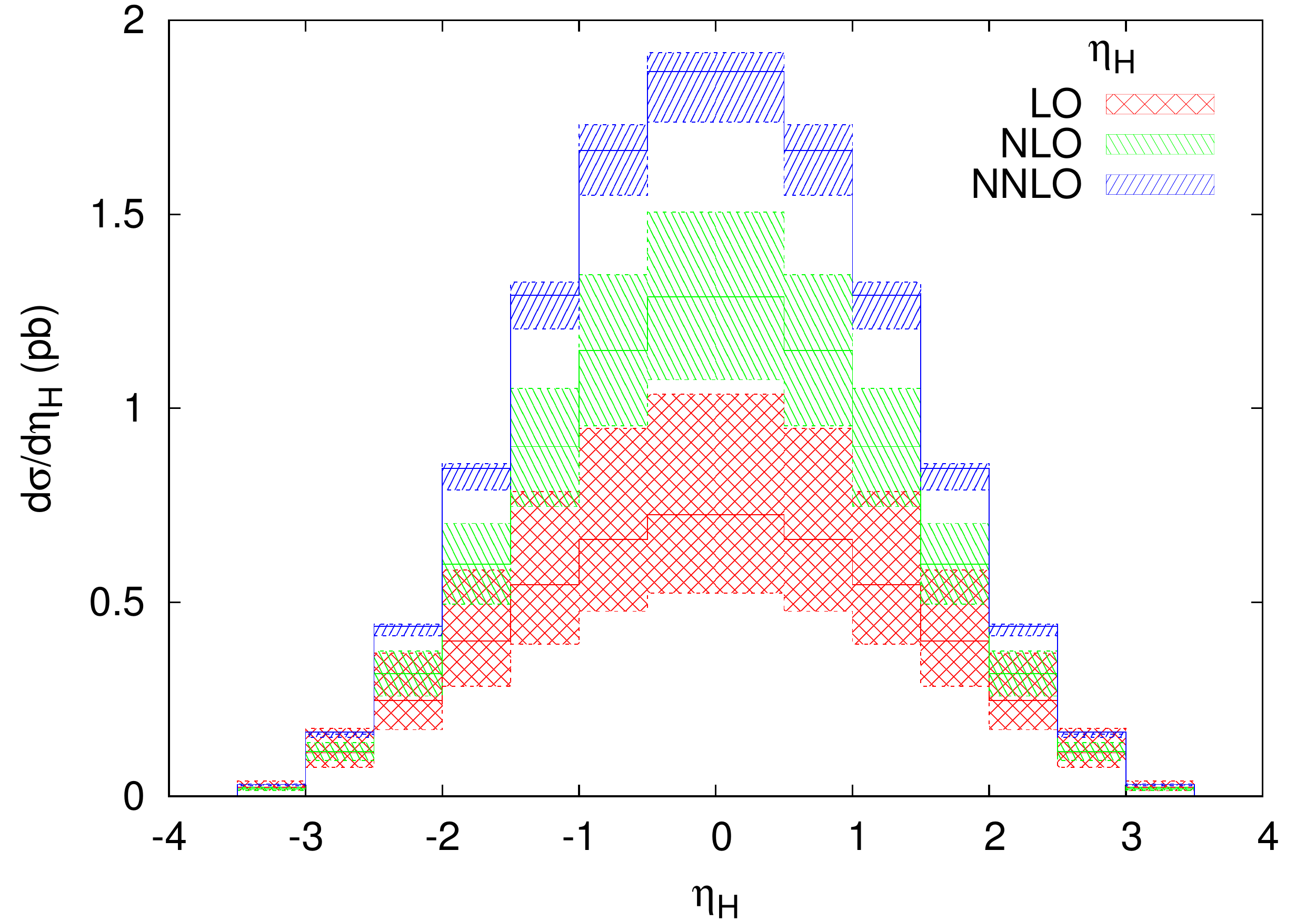}

(b)
\includegraphics[angle=-90,width=7cm]{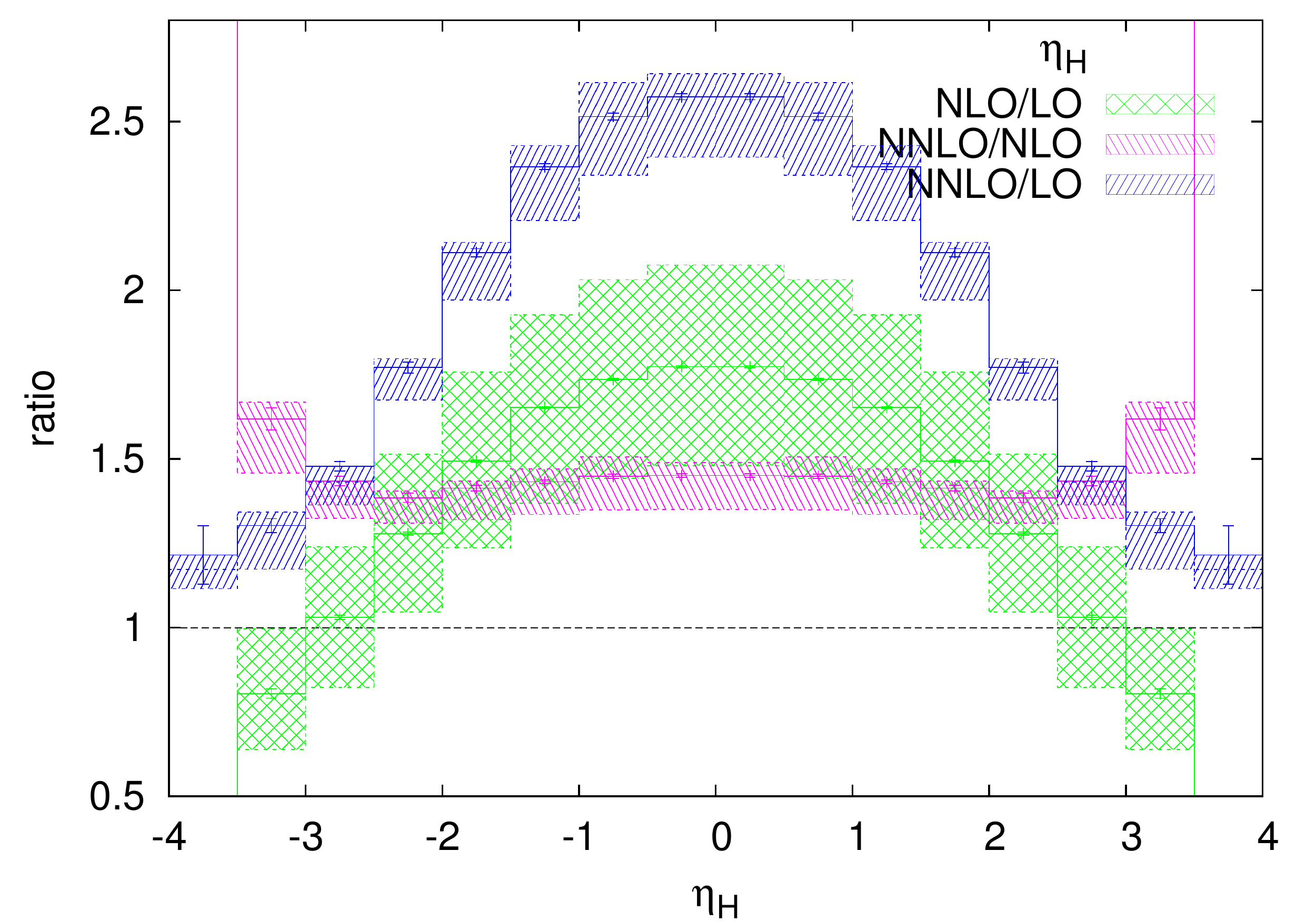}
\end{center}
\caption{(a) Rapidity distribution of the Higgs boson in inclusive $H+1j$ production 
 in $pp$ collisions with 
$\sqrt{s}=8$~TeV
at LO, NLO, NNLO and (b)
Ratios of different perturbative orders, NLO/LO, NNLO/LO and NNLO/NLO.\label{fig:etaH}}
\end{figure}

\begin{figure}[t]
\begin{center}
(a)
\includegraphics[angle=-90,width=7cm]{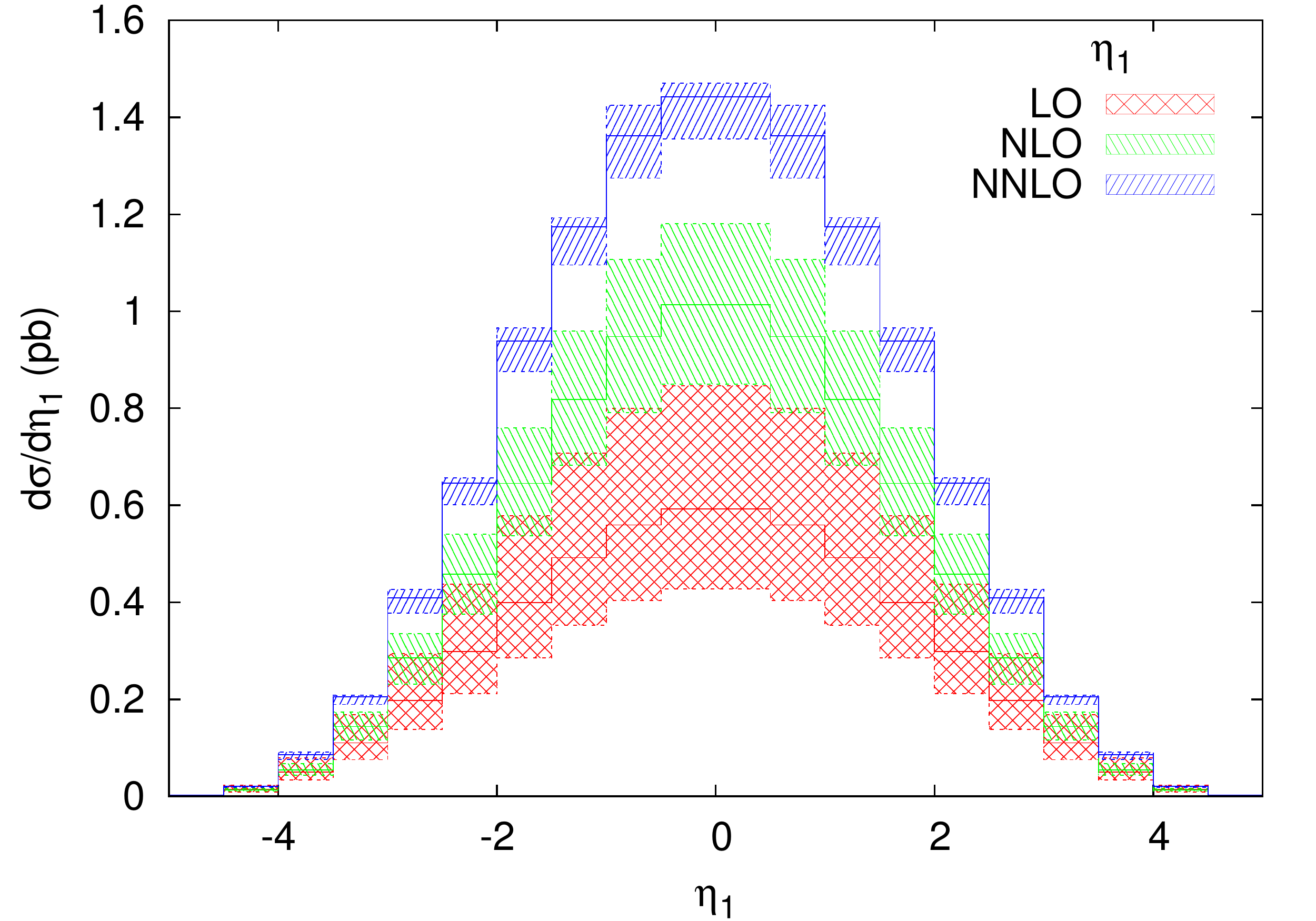}

(b)
\includegraphics[angle=-90,width=7cm]{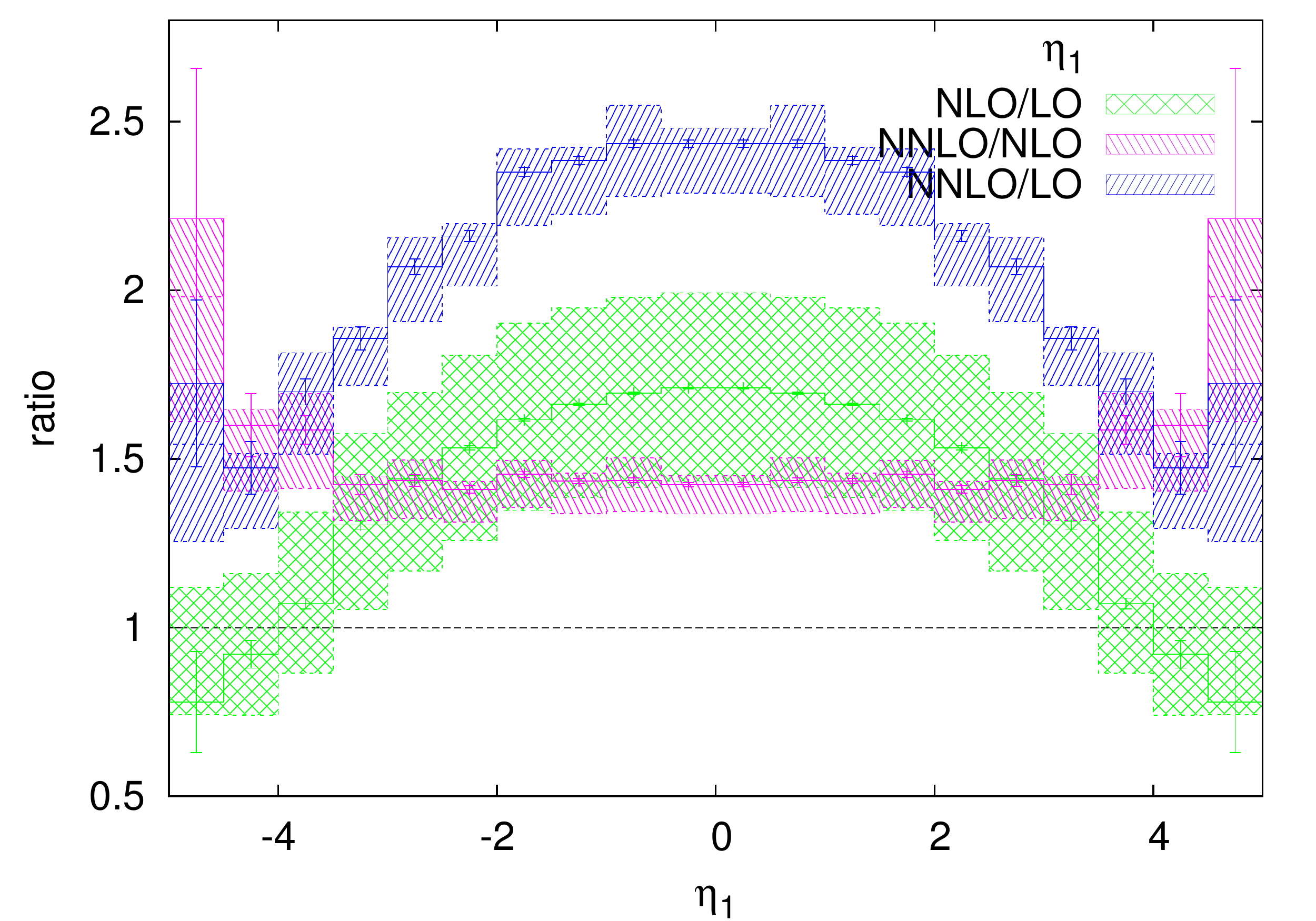}
\end{center}
\caption{
(a) Rapidity distribution of the leading jet in inclusive $H+1j$ production
 in $pp$ collisions with 
$\sqrt{s}=8$~TeV at LO, NLO, NNLO and (b)
ratios of different perturbative orders, NLO/LO, NNLO/LO and NNLO/NLO.\label{fig:etaJ}}
\end{figure}

The rapidity distribution of the Higgs boson and the leading jet are displayed in 
Figures~\ref{fig:etaH} and \ref{fig:etaJ} respectively. In both cases, we observe that the NLO 
corrections are largest at central 
rapidity, while becoming moderate at larger rapidities, while the ratio NNLO/NLO remains 
rather constant throughout the rapidity range. 
The residual theory uncertainty at NNLO is quasi constant for both distributions, and amounts to
9\%.
Both the transverse momentum 
and rapidity distributions highlight the fact that the NNLO corrections to $H+1j$ production 
in the gluon-only channel substantially enhance the normalization of NLO predictions, while 
not modifying the NLO shape,  except around the 
production threshold. 

In conclusion, we have described the first calculation of the fully differential
$H+1j$ cross sections at hadron colliders at NNLO in the
strong coupling constant using a new parton-level event generator. We have
considered the NNLO QCD corrections from the purely gluonic channel.  
Using the antenna subtraction scheme the explicit $\epsilon$-poles in the dimension regularization parameter of one- and two-loop matrix elements entering this calculation are cancelled in analytic and local form against the $\epsilon$-poles of the integrated antenna
subtraction terms thereby enabling the computation of Higgs plus jet cross sections at
hadron colliders at NNLO accuracy.  
The gluonic process yields the numerically largest contribution to 
$H+j$ final states, followed by the quark-gluon initiated and other subprocesses.   
However, the techniques employed here can be readily
applied to the quark contributions.  We observed that the 
transverse momentum and 
rapidity distributions of the 
Higgs boson and the leading jet receive 
substantial NNLO corrections. However, the shapes of the distributions do not change dramatically from 
NLO to NNLO, except around the production threshold in $p_T$. 

For all of the observables considered here, we observed a reduction of the respective
uncertainties in the theory prediction due to variations of the factorization
and renormalization scales, resulting 
in a residual uncertainty of around 9\% on the normalization of the 
distributions. We expect similar conclusions when including the
processes involving quarks. Our calculation brings $H+j$ production to the same level of 
theory accuracy as inclusive $H$ production, and will thus provide a crucial tool for
precision studies of the Higgs boson in the upcoming data taking periods at the CERN LHC. 

This research was supported in part by the Forschungskredit der Universit\"at Z\"urich, in part by
the Swiss National Science Foundation (SNF) under contract 200020-149517, in part by
the UK Science and Technology Facilities Council as well as by the Research Executive Agency (REA) of the European Union under the Grant Agreements PITN-GA—2010-264564 (``LHCPhenoNet''), PITN-GA–2012–316704  (``HiggsTools''), and the ERC Advanced Grant MC@NNLO (340983).


\begin{thebibliography}{99}

\bibitem{higgsLHC}
G.~Aad {\it et al.}  [ATLAS Collaboration],
  Phys.\ Lett.\ B {\bf 716} (2012) 1
  [arXiv:1207.7214];
  S.~Chatrchyan {\it et al.}  [CMS Collaboration],
  Phys.\ Lett.\ B {\bf 716} (2012) 30
  [arXiv:1207.7235].

\bibitem{higgspropsLHC}
G.~Aad {\it et al.}  [ATLAS Collaboration],
  ATLAS-CONF-2014-009;
  S.~Chatrchyan {\it et al.}  [CMS Collaboration],
  CMS-PAS-HIG-14-009; 
 V.~Khachatryan {\it et al.}  [CMS Collaboration],
  arXiv:1405.3455 [hep-ex].

\bibitem{ggHnnlo}
C.~Anastasiou, K.~Melnikov and F.~Petriello,
  Nucl.\ Phys.\ B {\bf 724} (2005) 197
  [hep-ph/0501130];
 M.~Grazzini,
  JHEP {\bf 0802} (2008) 043
  [arXiv:0801.3232].

\bibitem{vhnnlo}
 G.~Ferrera, M.~Grazzini and F.~Tramontano,
  Phys.\ Rev.\ Lett.\  {\bf 107} (2011) 152003
  [arXiv:1107.1164];
  JHEP {\bf 1404} (2014) 039
  [arXiv:1312.1669].

\bibitem{vbfnlo}
   T.~Figy, C.~Oleari and D.~Zeppenfeld,
  Phys.\ Rev.\ D {\bf 68} (2003) 073005
  [hep-ph/0306109];
 K.~Arnold, {\it et al.},
  Comput.\ Phys.\ Commun.\  {\bf 180} (2009) 1661
  [arXiv:0811.4559];
  F.~Campanario, T.M.~Figy, S.~Pl{\"a}tzer and M.~Sj{\"o}dahl,
  arXiv:1308.2932.
\bibitem{ttHnlo}
 W.~Beenakker, S.~Dittmaier, M.~Kramer, B.~Plumper, M.~Spira and P.M.~Zerwas,
  Nucl.\ Phys.\ B {\bf 653} (2003) 151
  [hep-ph/0211352];
  S.~Dawson, C.~Jackson, L.H.~Orr, L.~Reina and D.~Wackeroth,
  Phys.\ Rev.\ D {\bf 68} (2003) 034022
  [hep-ph/0305087];
  R.~Frederix, S.~Frixione, V.~Hirschi, F.~Maltoni, R.~Pittau and P.~Torrielli,
  Phys.\ Lett.\ B {\bf 701} (2011) 427
  [arXiv:1104.5613].
\bibitem{ggH1jnlo}
D.~de Florian, M.~Grazzini, Z.~Kunszt,
  Phys.\ Rev.\ Lett.\  {\bf 82} (1999)  5209
  [hep-ph/9902483];
 V.~Ravindran, J.~Smith, W.L.~Van Neerven,
  Nucl.\ Phys.\  {\bf B634} (2002)  247
  [hep-ph/0201114].

\bibitem{ggH2jnlo}
J.M.~Campbell, R.K.~Ellis, G.~Zanderighi,
  JHEP {\bf 0610} (2006) 028.
  [hep-ph/0608194];\\
 J.M.~Campbell, R.K.~Ellis, C.~Williams,
  Phys.\ Rev.\  {\bf D81} (2010)  074023.
  [arXiv:1001.4495].
 H.~van Deurzen, N.~Greiner, G.~Luisoni, P.~Mastrolia, E.~Mirabella, G.~Ossola, T.~Peraro and J.F.~von Soden-Fraunhofen {\it et al.},
  Phys.\ Lett.\ B {\bf 721} (2013) 74
  [arXiv:1301.0493].



\bibitem{ggH3jnlo}
 G.~Cullen, H.~van Deurzen, N.~Greiner, G.~Luisoni, P.\ Mastrolia, E.~Mirabella, G.~Ossola,
  T.~Peraro and F.\ Tramontano,
 Phys.\ Rev.\ Lett.\  {\bf 111} (2013) 131801
  [arXiv:1307.4737].
  
  \bibitem{ggHnew}
   C.~Anastasiou, C.~Duhr, F.~Dulat, E.~Furlan, T.~Gehrmann, F.~Herzog and B.~Mistlberger,
  arXiv:1403.4616.
  
 \bibitem{veto1} 
   S.~Catani, D.~de Florian and M.~Grazzini,
  JHEP {\bf 0201} (2002) 015
  [hep-ph/0111164].
  
  \bibitem{hcorr1}
   I.W.~Stewart and F.J.~Tackmann,
  Phys.\ Rev.\ D {\bf 85} (2012) 034011
  [arXiv:1107.2117].

  \bibitem{hcorr2}
   E.~Gerwick, T.~Plehn and S.~Schumann,
  Phys.\ Rev.\ Lett.\  {\bf 108} (2012) 032003
  [arXiv:1108.3335].
  \bibitem{caola}
   R.~Boughezal, F.~Caola, K.~Melnikov, F.~Petriello and M.~Schulze,
  JHEP {\bf 1306} (2013) 072
  [arXiv:1302.6216].
 
 \bibitem{h3g2l}
    T.~Gehrmann, M.~Jaquier, E.W.N.~Glover and A.~Koukoutsakis,
  JHEP {\bf 1202} (2012) 056
  [arXiv:1112.3554].
  
 \bibitem{h4g1l}
  L.J.~Dixon and Y.~Sofianatos,
  JHEP {\bf 0908} (2009) 058
  [arXiv:0906.0008];
  S.~Badger, E.W.N.\ Glover, P.~Mastrolia and C.~Williams,
  JHEP {\bf 1001} (2010) 036
  [arXiv:0909.4475]
  S.~Badger, J.M.~Campbell, R.K.~Ellis and C.~Williams,
  JHEP {\bf 0912} (2009) 035
  [arXiv:0910.4481].

 \bibitem{h5g0l}
  V.~Del Duca, A.~Frizzo and F.~Maltoni,
  JHEP {\bf 0405} (2004) 064
  [hep-ph/0404013];
  L.J.~Dixon, E.W.N.~Glover and V.V.~Khoze,
  JHEP {\bf 0412} (2004) 015
  [hep-th/0411092];
  S.D.~Badger, E.W.N.~Glover and V.V.~Khoze,
  JHEP {\bf 0503} (2005) 023
  [hep-th/0412275].
  
 \bibitem{secdec}
  T.~Binoth and G.~Heinrich,
  Nucl.\ Phys.\ B {\bf 693} (2004) 134
  [hep-ph/0402265];
  C.~Anastasiou, K.~Melnikov and F.~Petriello,
  Phys.\ Rev.\ D {\bf 69} (2004) 076010
  [hep-ph/0311311].
  
 \bibitem{ourant}
    A.~Gehrmann-De Ridder, T.~Gehrmann and E.W.N.\ Glover,
  JHEP {\bf 0509} (2005) 056
  [hep-ph/0505111];
  Phys.\ Lett.\ B {\bf 612} (2005) 49
  [hep-ph/0502110];
  Phys.\ Lett.\ B {\bf 612} (2005) 36
  [hep-ph/0501291].

 \bibitem{qtsub}
  S.~Catani and M.~Grazzini,
  Phys.\ Rev.\ Lett.\  {\bf 98} (2007) 222002
  [hep-ph/0703012].
  
 \bibitem{stripper}
  M.~Czakon,
  Phys.\ Lett.\ B {\bf 693} (2010) 259
  [arXiv:1005.0274].
 
 \bibitem{hadant}
  A.~Daleo, T.~Gehrmann and D.~Maitre,
  JHEP {\bf 0704} (2007) 016
  [hep-ph/0612257].
 
 \bibitem{gionata}
   A.~Daleo, A.~Gehrmann-De Ridder, T.~Gehrmann and G.~Luisoni,
  JHEP {\bf 1001} (2010) 118
  [arXiv:0912.0374].
 
 \bibitem{monni}
   T.~Gehrmann and P.F.~Monni,
  JHEP {\bf 1112} (2011) 049
  [arXiv:1107.4037].
 
 \bibitem{ritzmann}
  R.~Boughezal, A.~Gehrmann-De Ridder and M.~Ritzmann,
  JHEP {\bf 1102} (2011) 098
  [arXiv:1011.6631];
   A.~Gehrmann-De Ridder, T.~Gehrmann and M.~Ritzmann,
  JHEP {\bf 1210} (2012) 047
  [arXiv:1207.5779].
 
 \bibitem{joao}
  E.W.N.~Glover and J.~Pires,
  JHEP {\bf 1006} (2010) 096
  [arXiv:1003.2824];
   A.~Gehrmann-De Ridder, E.W.N.~Glover and J.~Pires,
  JHEP {\bf 1202} (2012) 141
  [arXiv:1112.3613];
  A.~Gehrmann-De Ridder, T.~Gehrmann, E.W.N.~Glover and J.~Pires,
  JHEP {\bf 1302} (2013) 026
  [arXiv:1211.2710];
  Phys.\ Rev.\ Lett.\  {\bf 110} (2013) 162003
  [arXiv:1301.7310].
  
   \bibitem{nnpdf}
   R.D.~Ball,  {\it et al.},
  Nucl.\ Phys.\ B {\bf 867} (2013) 244
  [arXiv:1207.1303].
  \bibitem{atlashqt}
   The ATLAS collaboration,
  ATLAS-CONF-2013-072.
  
  
  \bibitem{hqtresum}
   C.~Balazs and C.P.~Yuan,
  Phys.\ Lett.\ B {\bf 478} (2000) 192
  [hep-ph/0001103];
  G.~Bozzi, S.~Catani, D.~de Florian and M.~Grazzini,
  Nucl.\ Phys.\ B {\bf 737} (2006) 73
  [hep-ph/0508068];
  V.~Ahrens, T.~Becher, M.~Neubert and L.~L.~Yang,
  Eur.\ Phys.\ J.\ C {\bf 62} (2009) 333
  [arXiv:0809.4283];
   D.~de Florian, G.~Ferrera, M.~Grazzini and D.~Tommasini,
  JHEP {\bf 1111} (2011) 064
  [arXiv:1109.2109];
  JHEP {\bf 1206} (2012) 132
  [arXiv:1203.6321].
  
\end{thebibliography}
\end{document}